# X-ray Phase Contrast Tomography to assess the sequential evolution of multi-organ damage in an animal model of multiple sclerosis

F. Palermo[1,2], N. Pieroni[1], A. Sanna[1,2], B. Parodi[3], C. Venturi[4], G. Begani Provinciali[1], L. Massimi[1], L. Maugeri[2], E. Longo[5], L. D'Amico[5], G. Tromba[5], I. Bukreeva[1], M. Fratini[1], G. Gigli[2], Nicole Kerlero de Rosbo[3]* A. Cedola[1]*

[1] Institute of Nanotechnology- CNR, Rome, Italy  [2] Institute of Nanotechnology-CNR, Lecce, Italy  [3] Department of Neurosciences, Rehabilitation, Ophthalmology and Maternal-Fetal Medicine (DINOGMI), University of Genoa, Genoa, Italy  [4] IRCCS Ospedale Policlinico San Martino  [5] Elettra-Sincrotrone Trieste S.C.p.A., I-34149 Trieste (TS), Italy

*Equal contribution

## Abstract

We use X-ray phase-contrast tomography (XPCT) in a multi-organ approach to identify early imaging markers predictive of multiple sclerosis (MS) in EAE animal model. As the majority of neurodegenerative diseases, MS is characterized by a progressive accumulation of biological deficits across different organs and systems. A simultaneous imaging of different disease-relevant networks and a multiscale imaging, ranging from the single cell through to the organ as a whole, are required to provide complete reliable information. XPCT offers the unprecedented possibility to investigate structural and cellular alterations at brain, gut, and eye levels in a multiscale approach. The comparison between naive mice and EAE-affected mice sacrificed at different time points and the correlation of the data from different organs and different time points, unveils the identification of early changes in organs possible predictive of the disease.

## Introduction

Multiple sclerosis (MS) is an inflammatory demyelinating disease which results in a progressive damage to the structures of the central nervous system (CNS). It is the most common non-traumatic cause of neurological disability in young adults, but its etiology is still uncertain. MS is a complex and heterogeneous disease which presents variable clinical and pathological manifestations: inflammation process involving T cells, B cells, activated antigen presenting cells and complement; demyelination, as a consequence of myelin attacked immune system and/or death of oligodendrocytes, cells specialized in producing myelin; axonal and neuronal damage and loss.
As most neurodegenerative diseases, recent studies [Sweeney, M.D. et al., Blood–brain barrier breakdown in Alzheimer disease and other neurodegenerative disorders. Nature Reviews Neurology, 2018. 14(3): p. 133] suggest that MS might be characterized by a progressive accumulation of biological deficits in various systems and anatomical sites, such as intestine and eyes, besides CNS.
In the light of the so-called gut-brain axis - a bi-directional communication system between the central and the enteric nervous system - gut changes are suspected as potential inducers of several neurodegenerative diseases, with gut alterations shown to lead to brain dysfunction at multiple levels [Parodi, B., and Kerlero de Rosbo, N., "The Gut-Brain Axis in Multiple Sclerosis. Is Its Dysfunction a Pathological Trigger or a Consequence of the Disease?." *Frontiers in Immunology* (2021): 3911]. In MS as well as in its animal model, the experimental autoimmune encephalomyelitis (EAE), it has been reported that gut inflammation leads to intestinal barrier alteration and increased permeability [Buscarinu, M., et al., Intestinal permeability in relapsing-remitting multiple sclerosis. Neurotherapeutics, 2018. 15(1): p. 68-74. 57; Camara-Lemarroy, C.R., et al., The intestinal barrier in multiple sclerosis: implications for pathophysiology and therapeutics. Brain, 2018. 141(7): p. 1900-1916; Nouri, M., et al., Intestinal barrier dysfunction develops at the onset of experimental autoimmune encephalomyelitis and can be induced by adoptive transfer of auto-reactive T cells. PloS one, 2014. 9(9): p. e106335], although this matter is yet poorly described.

The eye is a unique window on the brain: the retina and the optic nerve originate as outgrowths of the developing brain, and they are considered to be part of CNS, rather than the peripheral nervous system. The demyelination induced by MS occurs in all myelin-rich tissues, including the optic nerve and the retinal nerve fibers as well as the brain and spinal cord. This makes the visual system an important location where to seek for imaging degeneration markers for MS.

Optic neuritis (ON), characterized by acute inflammation of the optic nerve, can be one of the first clinical manifestations of MS, but visual deficits can also occur in MS patients without a diagnosis of ON [Fisher, J. B. et al., Relation of visual function to retinal nerve fiber layer thickness in multiple sclerosis. Ophthalmology 113, 324–332 (2006); Monteiro, M. L., et al., Quantification of retinal neural loss in patients with neuromyelitis optica and multiple sclerosis with or without optic neuritis using fourier-domain optical coherence tomography. Invest. Ophthalmol. Vis. Sci. 53, 3959–3966 (2012)]. Almost all EAE mice immunized with MOG develop various degree of ON [Hui Shao, et al., Myelin/Oligodendrocyte Glycoprotein–Specific T-Cells Induce Severe Optic Neuritis in the C57Bl/6 Mouse, Investigative Ophthalmology & Visual Science, November 2004, Vol. 45, No. 11; Quinn, T.A., et al., Optic neuritis and retinal ganglion cell loss in a chronic murine model of multiple sclerosis, Frontiers in Neurology, Neuro-ophthalmology, Volume 2, Article 50-6 (2011); Soares, R.M.G., et al., "Optical neuritis induced by different concentrations of myelin oligodendrocyte glycoprotein presents different profiles of the inflammatory process." *Autoimmunity* 46.7 (2013): 480-485.], which manifests with inflammatory cell infiltrates and optic nerve atrophy [Larabee CM, et al., Myelin-specific Th17 cells induce severe relapsing optic neuritis with irreversible loss of retinal ganglion cells in C57BL/6 mice. Mol Vis. 2016 Apr 11; 22:332-41]. Since axons in the optic nerves are closely connected to the nerve fiber layer and ganglion cells [Jin J, et al., Glial pathology and retinal neurotoxicity in the anterior visual pathway in experimental autoimmune encephalomyelitis. Acta Neuropathol. Commun. 7(1):125. (2019)] degenerative processes involving the optic nerve might produce also retinal alterations, consisting in thinning of the nerve fiber layer and ganglion cell loss [Green, A. J., et al., Ocular pathology in multiple sclerosis: retinal atrophy and inflammation irrespective of disease duration. Brain 133, 1591–1601 (2010)].

Although vascular alterations play a relevant role in the progression of MS and EAE, the status of the vasculature at the different stages before and after the clinical onset has not yet been completely investigated. Studies of the CNS have described an evident loss of integrity of the vessels [Ge, Y., et al., Diminished visibility of cerebral venous vasculature in multiple sclerosis by susceptibility-weighted imaging at 3.0 Tesla. Journal of Magnetic Resonance Imaging: An Official Journal of the International Society for Magnetic Resonance in Medicine, 2009. 29(5): p. 1190-1194. 30; Kirk, S., et al., Angiogenesis in multiple sclerosis: is it good, bad or an epiphenomenon? Journal of the neurological sciences, 2004. 217(2): p. 125-130; Holley, J.E., et al., Increased blood vessel density and endothelial cell proliferation in multiple sclerosis cerebral white matter. Neuroscience letters, 2010. 470(1): p. 65-70] and an angiogenic remodeling significantly dependent on the time course of the disease [Roscoe, W., et al., VEGF and angiogenesis in acute and chronic MOG (35–55) peptide induced EAE. Journal of neuroimmunology, 2009. 209(1-2): p. 6-15; Boroujerdi, A., et al., Extensive vascular remodeling in the spinal cord of pre-symptomatic experimental autoimmune encephalomyelitis mice; increased vessel expression of fibronectin and the α5β1 integrin. Experimental neurology, 2013. 250: p. 43-51].

On the other hand, an X-ray phase-contrast tomography (XPCT) work [Cedola, A., et al., X-Ray Phase Contrast Tomography Reveals Early Vascular Alterations and Neuronal Loss in a Multiple Sclerosis Model. Sci Rep 7, 5890 (2017)] showed a decrease in vessel visibility at EAE onset as compared to healthy mice probably due to the formation of leaky angiogenic blood vessels, making them less detectable by XPCT, in accordance with the well-known loss of blood-brain barrier (BBB) integrity from early EAE stages [Roscoe, W., et al., VEGF and angiogenesis in acute and chronic MOG (35–55) peptide induced EAE. Journal of neuroimmunology, 2009. 209(1-2): p. 6-15]. BBB - and similarly the blood-spinal-cord-barrier (BSCB) - is a highly specialized structure that separates bloodstream cells from neural tissues with the function of protecting the CNS from toxic elements present in the blood circulation, while still allowing the passage of substances necessary for metabolic functions [Liebner, S., et al., Functional morphology of the blood–brain barrier in health and disease. Acta neuropathologica, 2018. 135(3): p. 311-336. 35; Daneman, R. and Prat, A., The blood–brain barrier. Cold Spring Harbor perspectives in biology, 2015. 7(1): p. a020412]. BBB/BSCB dysfunction is involved in many neurodegenerative diseases and has long been recognized as an important early feature of MS. During infiltration of immune cells and lesion formation, BBB/BSCB suffers of a dramatic increase of its permeability; function becomes compromised and vascular leakage with alteration at the junctional level occurs [Larochelle, C., et al., How do immune cells overcome the blood–brain barrier in multiple sclerosis? FEBS letters, 2011. 585(23): p. 3770-3780. 43; Greenwood, J., Mechanisms of blood-brain barrier breakdown. Neuroradiology, 1991. 33(2): p. 95-100.]. In EAE, the compromised barrier has been associated with infiltration of immune cells – such as macrophages, T and B cells – which extravasate into CNS (brain, spinal cord and optic nerve), because of their specificity for myelin antigens and undergo further activation [Stromnes, I.M., et al., Differential regulation of central nervous system autoimmunity by TH 1 and TH 17 cells. Nature medicine, 2008. 14(3): p. 337-342]. Magnetic resonance imaging showed that the first signs of BSCB disruption manifest before inflammation and demyelination, indicating that the increase in permeability precedes the destructive inflammatory process [Schellenberg, A.E., et al., Magnetic resonance imaging of blood–spinal cord barrier disruption in mice with experimental autoimmune encephalomyelitis. Magnetic Resonance in Medicine: An Official Journal of the International Society for Magnetic Resonance in Medicine, 2007. 58(2): p. 298-305]. Previous XPCT analyses on EAE tissues demonstrated that BBB/BSCB alterations can be directly detected at very early stage of the disease [Palermo, F., et al., X-ray Phase Contrast Tomography Serves Preclinical Investigation of Neurodegenerative Diseases, Frontiers in Neuroscience (2020): 1137]. XPCT provides an excellent non-invasive and highly sensitive 3D tool for the investigation of biological soft tissues, without sectioning, staining or exposing samples to aggressive tissue processing. XPCT enables a simultaneous imaging of different disease-relevant networks – such as the vascular and neural ones – and, mostly important, a multiscale imaging, ranging from the single cell through to the organ as a whole, with histology-like resolution. These properties prove to be crucial in the study of neurodegenerative diseases, where it is essential to gain knowledge about the interaction of the single element with the surrounding micro-environment.

Many of the problems associated with the BBB/BSCB are also highly pertinent to the blood-retinal barrier (BRB) [Ragelle, H., et al., Organ-On-A-Chip Technologies for Advanced Blood–Retinal Barrier Models. Journal of Ocular Pharmacology and Therapeutics, 2020. 36(1): p. 30-41], which controls immune cell migration from the bloodstream into the retina, together with the pigment epithelium. The BRB breakdown in EAE, which occurs with the degeneration of retinal ganglion cells and activation of glial cells [Schmitz, K., Tegeder, I., Bioluminescence and near-infrared imaging of optic neuritis and brain inflammation in the EAE model of multiple sclerosis in mice. J Vis Exp 121 (2017)], allows inflammatory cells to infiltrate directly into retina. As for many ocular disorders, BRB disruption may be involved in the pathogenesis of neurodegenerative diseases. [Shi, H., et al., Retinal capillary degeneration and blood-retinal barrier disruption in murine models of Alzheimer's disease, Acta neuropathologica communications 8.1 (2020): 1-20] In this frame, the investigation of the retinal vasculature could reveal the presence of pathological conditions manifesting through vascular alterations.

In the present work we present a methodological approach for XPCT multi-organ investigation, throughout the CNS, gut and eyes of EAE-affected mice, to identify early imaging indicators potentially acting as biomarkers of pathological processes. The data obtained in the different anatomical sites at several time-points were correlated to provide an assessment of the temporal progression of the disease.

Herein, we exploited XPCT to generate multiscale 3D images to reveal imaging alterations in the different anatomical sites, investigated at several pre-symptomatic stages until onset. BBB/BSCB permeability alteration was directly detected and monitored over time throughout the CNS. At gut level, we focused on ileum, which appears to be a preferential site for inflammation. In this region we revealed a variation of density of infiltrating cells in the lamina propria of villi. At eye level, we provided a 3D morphological description of optic nerve lesions due to EAE and demonstrated that XPCT is a feasible mean to detect the structure of blood vessels in eyes. We believe that a systematic study of eye vascularization may reveal pre-symptomatic structural alterations to be monitored during the neurodegeneration process associated with EAE/MS.

Such multi-organ analysis is intended to be the first step towards bringing together clinical and preclinical data to make predictions about the progression of the disease in the clinical settings. A detailed characterization of structural alterations in experimental model is essential, as it can facilitate translational research employing experimental designs that are directly adaptable to humans.

# Results

**XPCT allows direct detection of blood barrier disruption in the CNS.**
By means of XPCT we performed a 3D multi-organ investigation in order to uncover pre-symptomatic structural and cellular alterations due to EAE in CNS, gut and eye, monitoring the disease progression at early stages.

First of all, we investigated the disruption of BBB/BSCB, which is a crucial hallmark in the pathogenesis of MS and EAE. The impairment of the barrier manifests as an alteration in its permeability which favors subsequent leukocyte infiltration of immune cells and immune mediators into the CNS. We analysed brain and lumbar spinal cord (SC) of naïve mice as control and of EAE-affected mice. We focused on the lumbar tract of SC since it appears to be the initial site of EAE inflammation of CNS. The EAE mice were sacrificed at different time points, corresponding to 3 and 7 days after EAE induction (asymptomatic) and to the disease onset, occurring around 11$^{th}$ day post-immunization (dpi). Figure 1 demonstrates the progressive degeneration of the vascular system through the appearance of "clouds", visible as a halo around the vessels. These clouds – directly detected by means of XPCT - are compatible with extravasated material, consisting of inflammatory cells and blood proteins. The presence and the extension of the clouds can be assumed as a criterion to assess the increasing BBB/BSCB permeability and, consequently, the ongoing inflammatory process. Figures 1a-1b-1c-1d show representative tomographic images of a sample of lumbar SC in sagittal view for each time point. The tomographic images have a voxel size of $3.05 \times 3.05 \times 3.05$ µm$^3$. We applied image segmentation through intensity maximum projections in order to highlight the vasculature. The different grey-levels are proportional to different electron densities inside the sample: structures with higher electron density appear brighter than the surrounding tissue. We have chosen to use non-perfused samples to preserve the blood components inside them. In fact, thanks to the presence of proteins which bind and carry metals, such as hemoglobin, transferrin and albumin, blood acts as an endogenous contrast agent, highlighting the blood vessels. The micrometric spatial resolution allows the visualization of the distribution of vessels with diameters ranging from 10 to 35 microns.

In all the samples we can observe vessels arising from the longitudinal anterior spinal artery, visible as a white stripe running along the right vertical edge of the SC. In the lumbar SC of 7- and 11-day post injection (dpi) mice (Figures 1c-1d), these vessels appear surrounded by numerous clouds of an intermediate level of grey, which are not observed in the SC of naive mice or mice at 3 dpi (Figure 1a, b).

Quantification of total vascular damage at the different time points is shown in Figure 1e. We analyzed $n = 3$ samples for each group (naïve, 3 dpi, 7 dpi, EAE onset). The damage is expressed as the number of the detected vessels surrounded by clouds over the total number of visible vessels. At 3dpi, the first time point tested, damaged vessels are not visible. At 7 dpi, the second time point, about 10 % of the vessels were damaged; this value increased to 55% at disease onset
By measuring over time vascular damage separately for vessels originating from the central artery and for vessels from secondary arteries and venules, we wanted to investigate how vascular alteration spreads in lumbar SC. As mentioned before, at 7 dpi, vascular degeneration largely involves vessels arising from the anterior spinal artery, as shown in Fig1c and in Fig.1f, presenting the sagittal and axial views of a representative 7dpi lumbar SC respectively. The zoom in Fig 1f highlights the presence of a small cloud (indicated by the arrow) close to a vessel originating from the central artery. However, lesions are not visible around vessels arising from other arteries or venules. In contrast, at the onset of the disease, both vessels from the central artery and vessels from the minor arteries appear significantly damaged, as visible in Fig1d and in Fig.1g, showing the sagittal and axial views of a representative lumbar SC at the onset respectively. The zooms in Fig 1g highlights the presence of clouds around numerous vessels (arrows). Note that the clouds are mainly located at the base of the vessels.
Quantification of the number of damaged vessels measured separately for vessels originating from the central artery and for vessels from secondary arteries and venules, is reported in Fig1h. The vessels from the central artery appear to be almost all involved in the inflammatory process from 7dpi. On the other hand, for vessels originating from secondary arteries, we observe a sharp increase in damaged vessels from 15% at 7dpi to 50% at the onset, reflecting how the lesion becomes more intense and extensive at this stage.

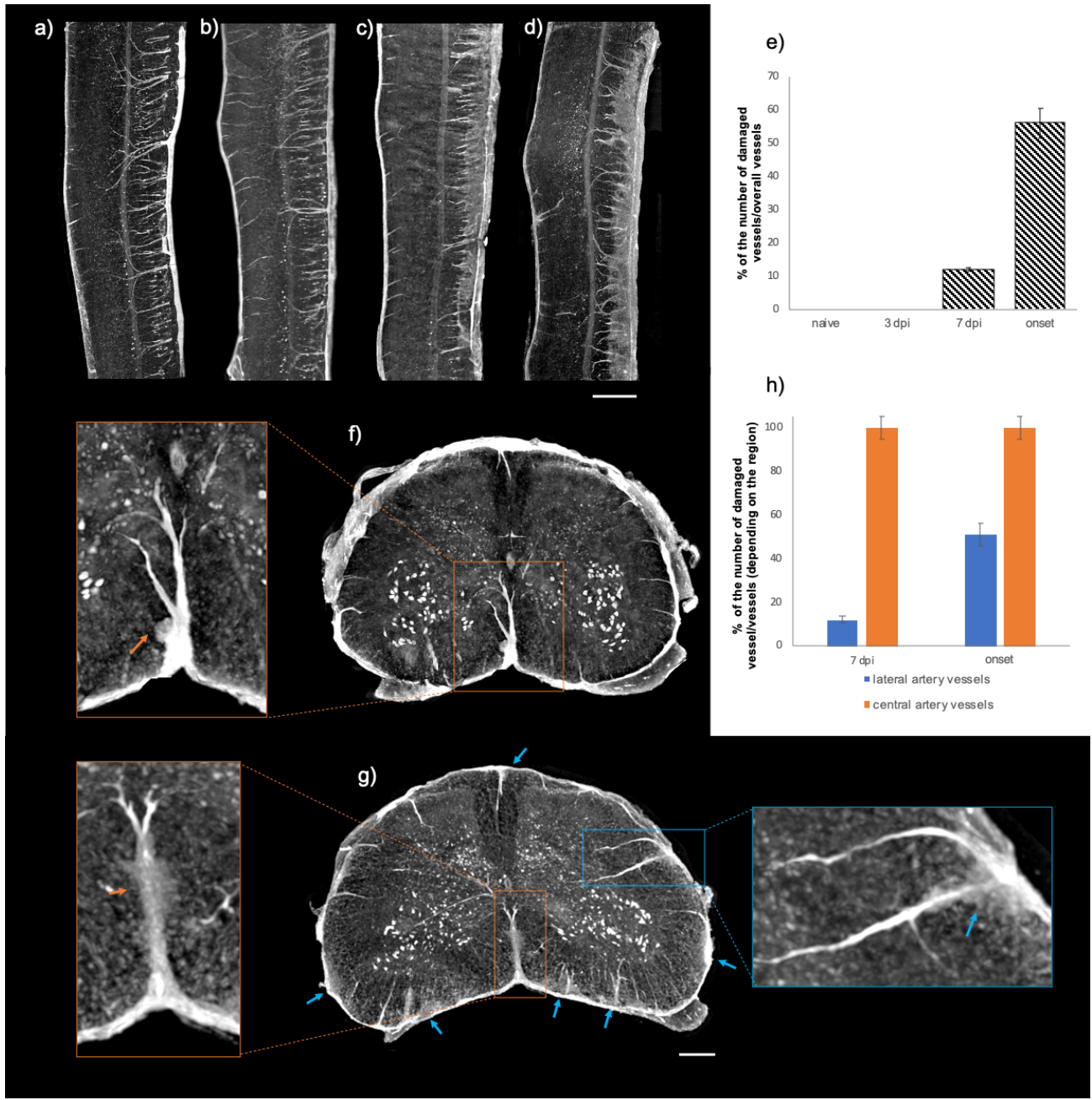

**Figure 1. XPCT imaging and quantification of BSCB leakage and lesion in lumbar spinal cord of EAE-affected mouse at early stages of the disease.**
**a-d)** XPCT images showing the sagittal views of the lumbar spinal cord in a naïve mouse (a) and EAE-injected mouse at 3 dpi (b), 7 dpi (c) and at the disease onset (d). In c) and d) the vessels arising from the anterior spinal artery appear surrounded by numerous clouds of extravasated material reflecting the intense BBB dysfunction in the EAE-affected mouse. Scale bar = 500 micron. **e)** Quantification of the number of damaged vessels with respect to the total number of vessels in lumbar spinal cord samples of EAE-affected mice (n = 3 per group). The data are presented as mean ± SEM. **f-g)** XPCT images showing the axial view of lumbar spinal cord of an EAE-injected mouse at 7 dpi (f) and at the onset (g). The insets and the arrows highlight the presence and the localization of clouds. At 7 dpi (f), vascular degeneration appears to involve vessels arising from the anterior spinal artery (inset). In contrast, at the onset (g) the lesion becomes more extensive, involving also vessels arising from minor arteries (arrows and blue inset). Scale bar = 200 micron. **h)** Quantification of the number of damaged vessels over the total number of vessels, separately for vessels originating from the central artery (orange) and for vessels from secondary arteries and venules (blue), in lumbar spinal cord samples from EAE mice at 7dpi and at the clinical onset (n = 3 per group). The data are presented as mean ± SEM. Images a-b-c-d-f-g- were obtained as maximum intensity projections XPCT volumes (a,b,c,d: over 150 micron; f,g: over 100 micron).
XPCT images were acquired in Experiment 1, reported in Methods.

Higher resolution images (voxel size of 0.65×0.65×0.65 µm$^3$) permit to visualize the clouds as a large accumulation of cells localized around the vessels, as seen in Figure 2a, which would be typical of an EAE lesion with infiltrating inflammatory T-cells and macrophages. The XPCT image shows the coronal view of lumbar SC of an EAE-affected mouse at the onset. The virtual rendering reported in Figure 2b provides a 3D morphological description of the surroundings of the vascular lesion enclosed in the white rectangle in Figure 2a. The structures of interest are segmented in different colors: we can observe how the small cells (gray) that compose the cloud completely surround the base of the vessels (yellow). As for the neural network, we can distinguish nonspecific neuron-like cells and multipolar neuron-like cells, rendered in green and blue respectively.

The SC tomographic images displayed in Figure 2c-2e demonstrate the level of detail achievable with micro-XPCT, which allows a thorough imaging of the micro-environment at neuronal and vascular level. Figure 2c shows cells located in the anterior horn of spinal cord compatible with neurons provided with processes, dendrites or axons. As observed in figures 2d-2f, we can detect dashed structures, highlighted by white arrows, which could be compatible with neural axons wrapped by myelin sheath, an extension of oligodendrocyte plasma membrane, periodically interrupted by the so-called nodes of Ranvier. To note, in Figure 2d, the myelinated axon (white arrow) seems to become less bright and to lose its periodicity as it approaches the EAE vascular lesion (black arrow). Figures 2e and 2f show a tomographic image and the same volume rendered in 3D, respectively, of a cell with dashed processes (arrows), compatible with a neuron with myelinated axons. In the 3D rendering the neuron and axons are segmented in red, while neuron-like cells in the surrounding are rendered in green. The 3D rendering of the volume allows to better visualize the spatial development of the axons.

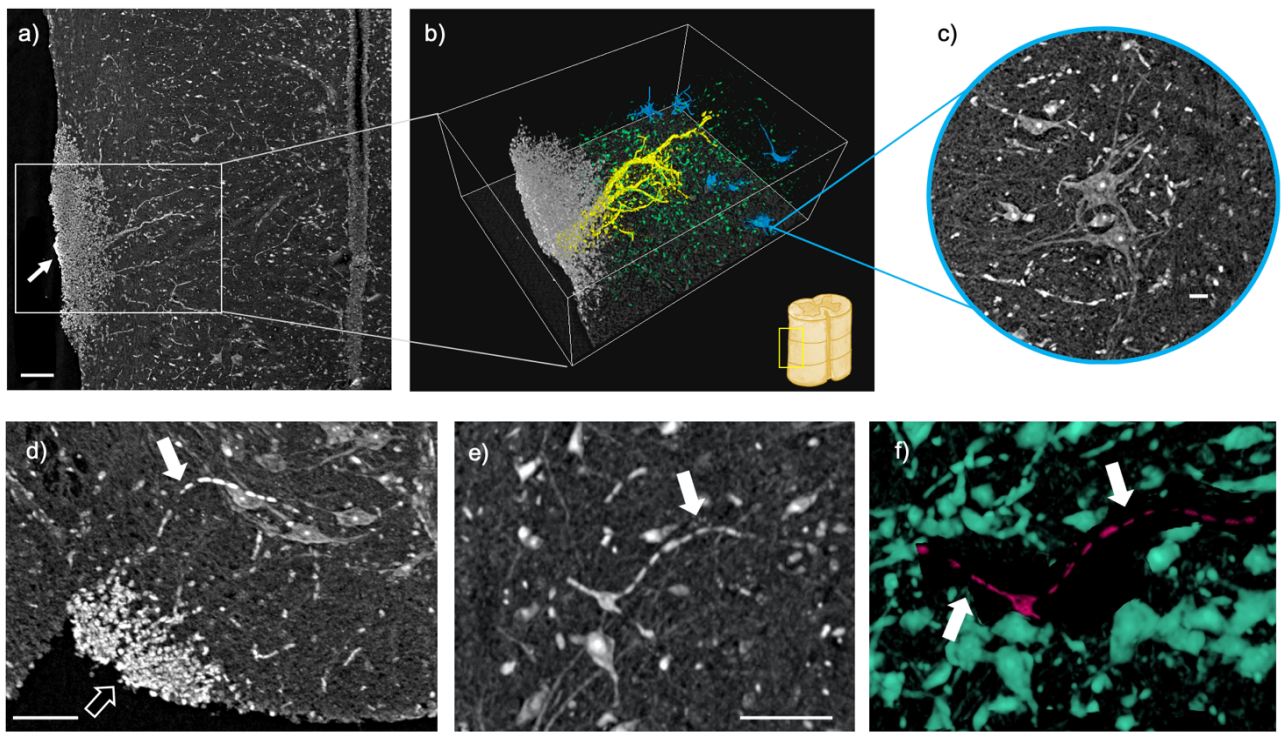

**Figure 2. High resolved XPCT describes the micro-environment of EAE vascular lesion in lumbar SC.**
**a)** XPCT image showing a portion of the lumbar SC of an EAE mouse at the disease onset in coronal view. The arrow highlights the presence of a large accumulation of cells around vessels arising from SC wall. Scale bar = 50 micron. **b)** 3D rendering of the micro-environment around the vascular lesion. The features are segmented with different colors: in white the small cells surrounding the vessel (yellow), in green and blue morphologically neuron-like cells. Inset: schematic representation of the localization of the vascular lesion of (a) and (b) in the SC. **c)** XPCT image of star-shaped cells located in the anterior horn of EAE mouse, compatible with neurons provided with processes, dendrites or axons. Scale bar = 25 micron. **d-e)** XPCT images with well-visible dashed structures (white arrows) compatible with axons wrapped by myelin sheath (scale bar = 50 micron). In (d), near the lesion (black arrow) the dashed structure seems to lose its periodicity and to become less bright. **f)** XPCT 3D-rendering of the region reported in (e): a cell with dashed processes is segmented in red, cells compatible with neurons in the surrounding are rendered in green. The 3D view of the volume allows to better visualize the spatial development of the neuronal processes.
Images a-c-d-e were obtained as maximum intensity projections XPCT volumes (a: over 15 micron; c: over 10 micron; d, e: over 10 micron).
XPCT images were acquired in Experiment 2, reported in Methods.

Along with the SCs, we measured brains dissected from the same mice, to assess the anatomical progression of the vascular damage throughout the CNS of EAE-affected mice. In Figure 3a is reported the posterior brain region of a mouse at 11 dpi, where cerebellum, brainstem and cervical SC regions are visible. Tomographic images in Figure 3b and figure 3c, which represent a portion of brainstem and cervical SC in coronal view respectively, demonstrate the presence of clouds and cell accumulation (highlighted by arrows) in these sites at the clinical onset. Although present with less occurrence with respect to the lumbar SC region, BBB/BSCB damage appears to have reached upper SC and brainstem at the clinical onset, while vascular alterations in these regions are not observed at 7 dpi. These results confirm that the vascular degeneration proceeds from the lumbar spinal cord to the brain, allowing the assessment of the temporal course of the BBB dysfunction.

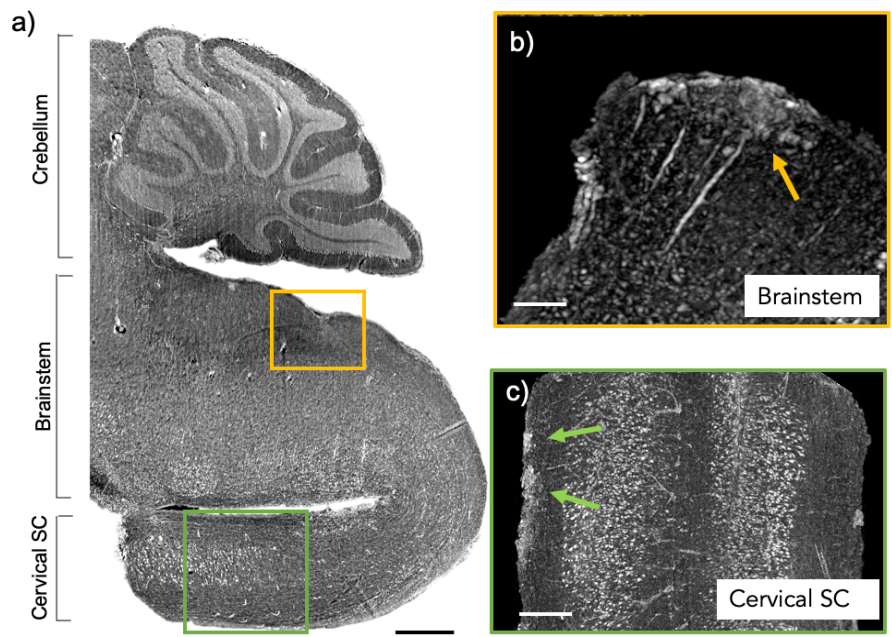

**Figure 3. XPCT shows vascular damage in EAE brain at clinical onset.**
**a)** XPCT image showing a representative sagittal view of cerebellum, brain stem and cervical SC of an EAE-affected mouse at disease onset. The squares indicate where the lesions showed in (b) and (c) are located. Scale bar = 500 micron. **b)** XPCT image of a detail of the brainstem at the EAE onset with the presence of clouds (indicated by the yellow arrow) at the base of the vessels (scale bar = 50 micron). **c)** XPCT image showing coronal view of the cervical SC of EAE-affected mouse at the onset. Green arrows highlight the vascular lesions (scale bar = 150 micron).
Images were obtained as maximum intensity projections XPCT volumes (a: over 150 micron; b: over 80 micron; c: over 100 micron).
XPCT images were acquired in Experiment 3, reported in Methods.

**Quantification of cell density in the gut.**
To investigate whether there is a temporal correlation between CNS and gut alterations, we sought for imaging markers for the inflammation processes which involve the intestine of EAE-affected mice. To reveal changes between normal and pathological conditions, the gut morphology needs to be examined, but gut complex geometry and structural convolutions prevent from identifying sample features. Anatomically, the gut looks like a long tube of varying diameter with flexible walls, folded several times on itself, as illustrated by 3D rendering in Figure 4a. To extract morphological details, we reduced the geometrical complexity by virtual flattening. The gut cylindrical surface was mapped into a plane by means of flattening algorithms described in Methods. These procedures allow to visualize structures otherwise very hard to recognize in the 3D volume. Figures 4b – 4c – 4d illustrate some morphological features of the small intestine wall, displayed on 2D plane upon virtual flattening. We can distinguish the thin longitudinal layer of smooth muscle fibers belonging to the tunica muscularis (Fig. 4b), cells compatible in shape, dimensions and location with neurons (Fig. 4c) of the myenteric plexus and blood vessel (Fig.4d) running along the tela submucosa.

XPCT do make possible the multiscale 3D imaging of the tissues, ranging from the intestine as a whole (Figure 4a) down to the single cell (Fig. 4b-c), overcoming the limitations of histology/immuno-histochemistry. They are well-consolidated and widely used techniques which provide very informative and resolved 2D images of biological tissues, but they require destructive sample preparation to thin the sample down to hundreds of microns and subsequently restrict spatial coverage within a finite depth. However, XPCT and histology/immuno-histochemistry are complementary techniques: the latter provides specific information to identify cells and other components, while XPCT provides 3D images but it cannot identify a cell other than through morphological aspect. Nevertheless, XPCT can reproduce in 3D morphological features visible in the histological sections with a similar spatial resolution. The comparison between Figure 4e and Figure 4f, showing a hematoxylin-eosin stain histological section and an XPCT image of villi from an EAE mouse ileum respectively, demonstrates indeed that XPCT can achieve histology-like resolution.

In some models of EAE, intestinal barrier permeability, together with morphological alterations, was observed already at 7 dpi and was associated with an increase in potentially pathogenic T cells infiltrating the gut lamina propria [Nouri M, et al., Intestinal Barrier Dysfunction Develops at the Onset of Experimental Autoimmune Encephalomyelitis, and Can Be Induced by Adoptive Transfer of Auto-Reactive T Cells. PloS One (2014) 9(9):e106335]. In order to identify an imaging marker of EAE detectable with XPCT, we measured the density of cells located in the lamina propria, limited to the region of the villi, at different time points in the disease (3dpi, 7 dpi, EAE onset, 4 days post EAE onset). A 3D rendering from XPCT data of an ileum villus is shown in Figure 4g, where blood vessels are segmented in red, lymphatic canals in blue and cell nuclei in yellow. By means of XPCT indeed, we are not able to identify a cell, if not for morphological or location considerations. For this reason, we quantified the variation of the number of cells localized in the area delineated by the red line (excluding thus the epithelial cells), visible in Figure 4h, which reports a 3D rendering of a villus segmented to show the spatial distribution of cell nuclei (in yellow) inside the volume.

Cell density was calculated as the ratio of the number of cells in the outlined volume (Fig 4h) of the villus to the volume itself. We then normalized the cell density at the different time points to the value measured for the naïve mice. We analyzed $n = 2$ ilea for each time point (naïve, 3dpi, 7 dpi, EAE onset, EAE 4dpo). Measurements of cell density were performed on approximately 20 villi per ileum Quantification of cell density is shown in Figure 4i, where we observe a significant increment of density at the clinical onset. At a later stage of the disease (EAE 4 dpo) the density is reduced.

However, because of the large variability we have detected in the villi population of a single mouse and among different mice, we do not pretend to have obtained determinative or conclusive results. To achieve statistical significance, it is necessary to measure many more samples so as to limit the variability effect. Through this analysis we want to demonstrate that XPCT and the described procedures are fully suitable to perform both qualitative and quantitative investigations.

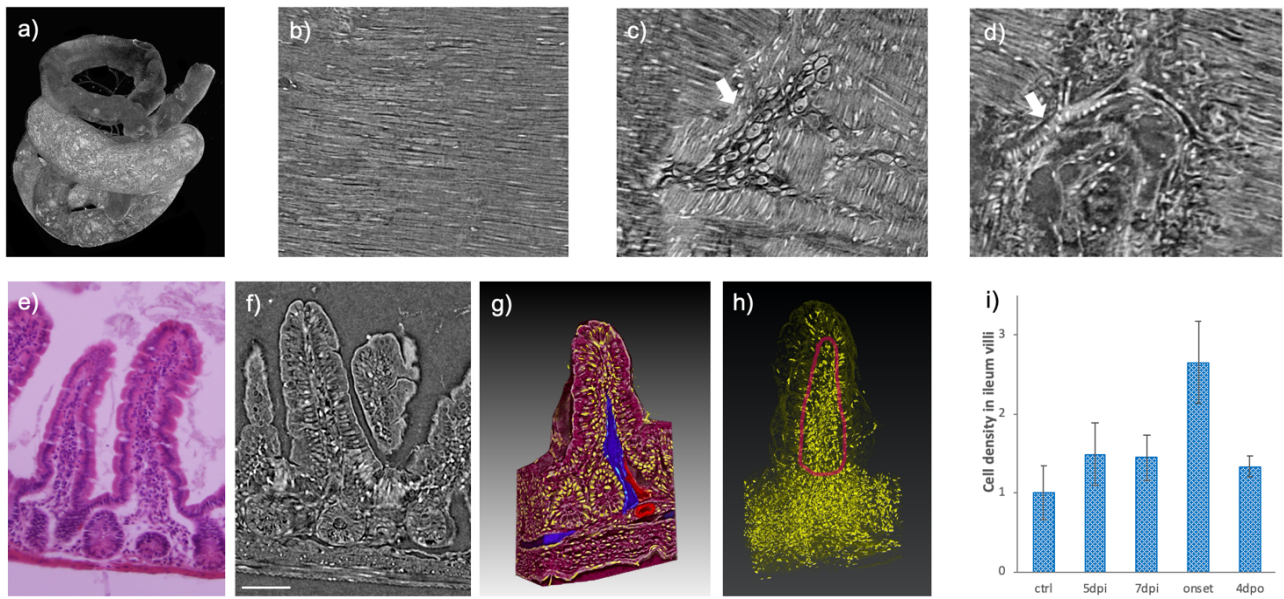

**Figure 4. XPCT multiscale imaging of mouse gut with histology-like resolution and quantification of cell density in lamina propria of ileum villi.**
**a)** 3D rendering of a whole mouse gut imaged with micro-XPCT. **b-c-d)** XPCT images of the layers of gut wall obtained after virtual flattening of the ileum: (b) Tunica muscolaris; (c) cells compatible in size, dimensions, location with neurons from Myenteric plexus (arrow) [Sibaev, A., et al. "Nociceptin effect on intestinal motility depends on opioid-receptor like-1 receptors and nitric oxide synthase co-localization." *World Journal of Gastrointestinal Pharmacology and Therapeutics* 6.3 (2015): 73; Wang, Hongtao, et al. "The timing and location of glial cell line-derived neurotrophic factor expression determine enteric nervous system structure and function." *Journal of Neuroscience* 30.4 (2010): 1523-1538.]; (d) vessel (arrow) running along the tela submucosa. **e)** Hematoxylin-eosin stain histological section of ileum villi of an EAE mouse. **f)** XPCT image of ileum villi of an EAE mouse demonstrates that this technique can achieve histology-like resolution. The image was obtained as maximum intensity projections over 5 micron. Scale bar = 50 micron. **g)** 3D rendering form XPCT data of ileum villus, where blood vessels are segmented in red, lymphatic canals in blue and cells in yellow. **h)** 3D rendering of the cells inside the volume of an ileum villus; the area outlined by the red line roughly corresponds to the region of the lamina propria. **i)** Quantification of cell density calculated as the ratio of the number of cells in the lamina propria to the volume occupied by this structure in the villus.
XPCT data were acquired in Experiment 2, reported in Methods.

**Eye and Optic Nerve**
Visual deficits are relevant symptoms in MS and EAE. Neuropathological alterations in the retina and the optic nerve occur: thinning of retinal fiber layers, losses in retinal ganglion cells, demyelination and activation of microglia and astroglia, inflammatory cell infiltration [Manogaran P, et al., (2019) Retinal pathology in experimental optic neuritis is characterized by retrograde degeneration and gliosis. Acta Neuropathol Commun 7(1):116; London, A., et al., The retina as a window to the brain—from eye research to CNS disorders. Nature Reviews Neurology, 9(1), 44-53 (2013)].
The accessibility of the retina with advanced non-invasive ocular imaging techniques makes it a potential convenient site for the research and the diagnosis of diseases affecting the CNS. For this reason, it is essential to identify imaging markers in the eye and to temporally correlate them to the progression of the disease in the other anatomical sites.
ON, which involves retinal ganglion cell apoptosis, inflammatory cell infiltration and demyelination of the optic nerve, often manifests in MS and EAE. Figures 5a-5b show XPCT images in longitudinal view of the optic nerve, near the point where it enters the retina, partially visible at the top of the images, in a naïve mouse and an EAE-affected mouse respectively. The dashed lines illustrate the location of the axial sections reported in Figures 5c-5d. In both longitudinal and axial view of EAE optic nerve (Fig. 5b and 5d), XPCT reveals clear alterations in this region as large accumulations of inflammatory cell infiltrates (highlighted by the arrows) compatible with the occurrence of ON. Moreover, the longitudinal view in Fig 5b clearly shows atrophy of the optic nerve.

XPCT technique provides a feasible mean to detect the structure of blood vessels, and we are interested in studying the vasculature of the eye over an area as large as possible. Blood vessels in the eye are confined mainly within very thin spherical layers; optimal visualization of vascular network is possible only through virtual flattening. Moreover, deformations occur upon the embedding media, clearly visible in Figure 5e, which shows a section of mouse eye imaged by XPCT. The outer layers of the eye collapsed onto the lens (the circular shape structure in the center) and curled up, since the vitreous contained within the eye dried out. Therefore, to investigate the ocular vascularization virtual flattening procedures are mandatory. Once the spherical surface of the eye has been virtually flattened, the different layers can be studied by simply performing virtual slicing parallel to the plane of the layers. Representative results of virtual flattening, performed by means of algorithms described in Methods, are shown in Figures 5f and 5g, which show a lateral section of the retina and the en-face surface of the choroidal layer, respectively. The choroid (Figure 5g) is a highly vascularized structure which separates the sclera from the retina. XPCT images were quantified and we measured the diameter of the visible vessels, which ranged from 5 to 20 micron.

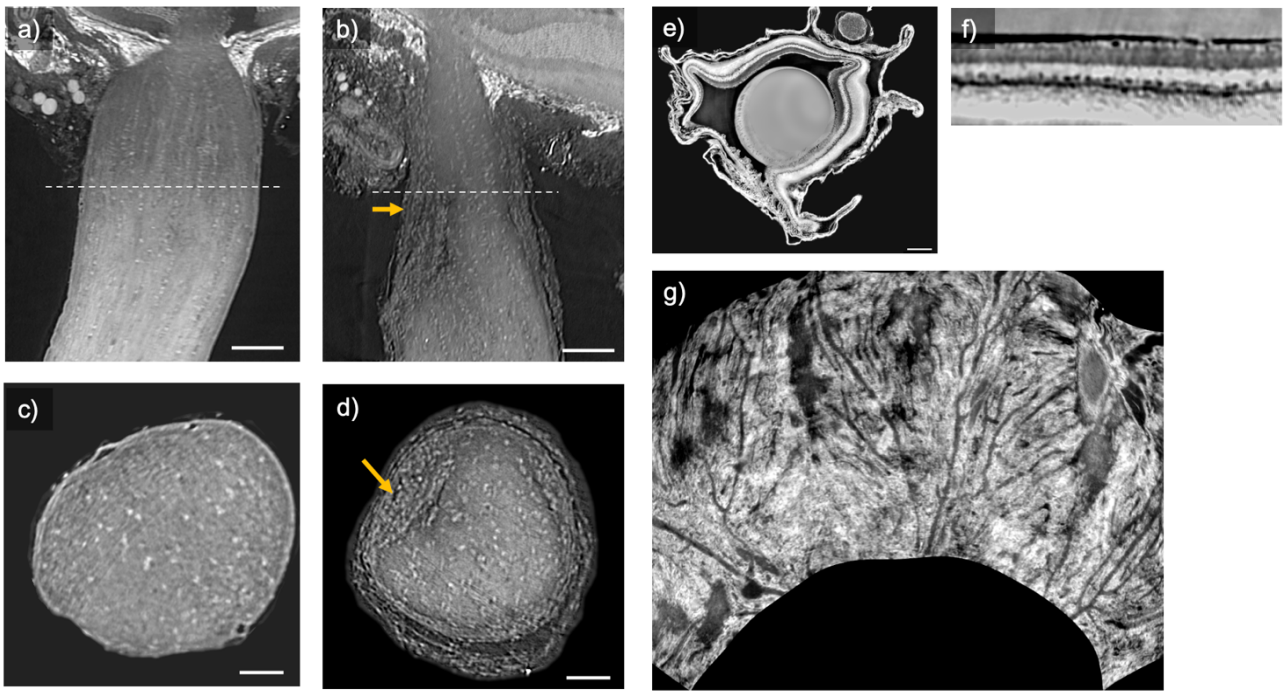

**Figure 5. XPCT shows pathological alterations in EAE optic nerve and allows to visualize ocular vascularization. a-d)** XPCT images showing the longitudinal (a, b) and axial (c, d) views of the optic nerve in a naive mouse (a, c), and an EAE-affected mouse at disease onset (b, d), where large accumulations of inflammatory cell infiltrates (arrows) compatible with the occurrence of optical neuritis is visible. The dashed lines in (a, b) illustrate the location of the axial sections (c, d). Scale bar a, b = 100 micron. Scale bar c, d = 50 micron. **e)** XPCT image showing a section of ocular bulb from a naïve mouse. Structural deformations due to sample preparation are visible. Scale bar = 200 micron. **f-g)**

Representative results of virtual flattening: lateral section of retina (f) and the en-face surface of the highly vascularized choroidal layer (g).
XPCT images were acquired in Experiment 4, reported in Methods.

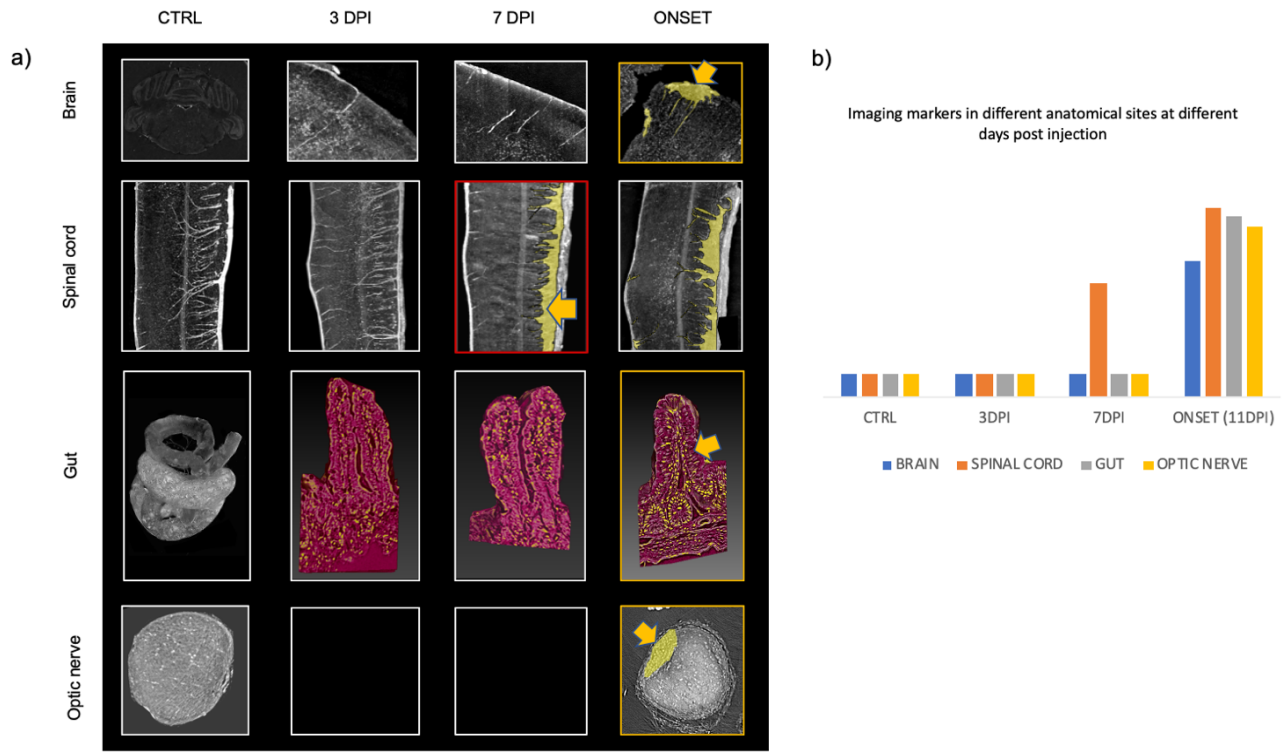

**Figure 6. Multi-organ investigation and temporal assessment of EAE imaging markers at early stages.**
**a-b)** Schematic figures summarizing the multi-organ investigation at different time-points of EAE imaging markers. The arrows highlight the presence of alterations detected by XPCT. Degeneration initially appears in the SC at 7 dpi, where it emerges as an alteration of the vascular network. It subsequently spreads to the brain, intestine and optic nerve, particularly at the onset of the disease. In the brain the degenerative process involves vascularization as well as in the SC. In the intestine and in the optic nerve, on the other hand, the disease manifests through an increase in cellular infiltrates.

## Methods

**EAE induction and sample preparation.**
Female C57BL/6 J mice, 6 to 8 weeks old, weighing 18.5 g purchased from Harlan Italy, were immunized as described before [Mendel, I., et al., A myelin oligodendrocyte glycoprotein peptide induces typical chronic experimental autoimmune encephalomyelitis in H-2b mice: Fine specificity and T cell receptor Vβ expression of encephalitogenic T cells. European journal of immunology, 1995. 25(7): p. 1951-1959] by subcutaneous injection (200 ml total) at two sites in the flank with an emulsion of 200 mg myelin oligodendrocyte glycoprotein (MOG) peptide 35–55 (Espikem) in incomplete Freund adjuvant (Difco) containing 600 mg Mycobacterium tuberculosis (strain H37Ra; Difco). The mouse was injected (100 ml total) in the tail vein with 400 ng pertussis toxin (Sigma-Aldrich) immediately and 48 h after immunization. The mice were scored daily for clinical manifestations of EAE on a scale of 0–5 [Mendel et al., 1995], and sacrificed by $CO_2$ inhalation at onset of clinical manifestations (around day 11 after immunization), with a clinical score of 3.5.

Samples consist of brain, lumbar/sacrococcygeal (L1–S4) spinal cord, gut and eyes, dissected out from mice sacrificed at several time-points. Mice were not perfused to preserve the blood inside the vessels [Frontiers]. Further preparation steps were required depending on the experimental conditions under which they were to be measured.

All animals are housed in pathogen-free conditions and treated according to the Italian and European guidelines (Decreto Legislativo 4 marzo 2014, n. 26, legislative transposition of Directive 2010/63/EU of the European Parliament and of the Council of 22 September 2010 on the protection of animals used for scientific purposes), with food and water ad libitum. The research protocol was approved by the Ethical Committee for Animal Experimentation of the University of Genoa (Prot. 319).

**Preparation for CNS samples for XPCT experiment 1.** Lumbar/sacrococcygeal (L1–S4) spinal cord from naive and EAE-affected mice at 3dpi, 7 dpi and 11 dpi ($n = 3$ for each group) were fixed in 4% paraformaldehyde for 24 h, then stored in 70% ethanol until XPCT. Just before the measurements, the samples were embedded in agarose gel, which keeps them hydrated and prevent radiation damage and movement during the experiment.

**Preparation for CNS and gut samples for XPCT experiment 2 and 3.** Lumbar/sacrococcygeal (L1–S4) spinal cords, brains, ileum and colon samples from naive and EAE-affected mice at 3dpi, 7 dpi and at the clinical onset ($n = 2$ for each group) were dehydrated through a graded ethanol series (70/95/100%), put in propylene oxide, and included in paraffin.

**Preparation for eye samples for XPCT experiment 4.** Eyes with optic nerve from naive and EAE-affected mice at the onset were embedded in paraffin and in epon. $n = 2$ samples for each group (naïve and EAE) were dehydrated through a graded ethanol series (70/95/100%), put in propylene oxide, and included in paraffin. The remaining samples, also in this case n=2 for each group, were instead fixed in glutaraldehyde (2.5% in 0.1 M cacodylate buffer, pH 7.4 TA; 2 h), postfixed in osmium (1% in 0.1 M cacodylate buffer, pH 7.4; 2 h) and uranyl acetate (1% in water; overnight). Samples were then dehydrated through a graded ethanol series (70/96/100%), put in propylene oxide, and embedded in resin (Propylene+EPON) at 42°C overnight and for 48 hours at 60°C. It was found that epon preparation helps to better preserve the integrity of the eye shape.

**Experimental set-ups**

The XPCT experiments were performed (1) at the medical beamline ID17 of the European Synchrotron Radiation Facility (ESRF, Grenoble, France), (2) at the ANATOMIX beamline of Synchrotron SOLEIL (Paris, France), (3) at I13-2 beamline of Diamond Light Source (Didcot, UK) and (4) at SYRMEP beamline of Elettra Synchrotron (Trieste, Italy) in free-space propagation mode [Bravin, A., et al., X-ray phase-contrast imaging: from pre-clinical applications towards clinics. Physics in Medicine & Biology, 2012. 58(1): p. R1].

(1) Data acquisition at ID17 was carried out using monochromatic incident X-ray beam with an energy of 35 keV. The sample-detector distance was set at 2.3 m. The detector had an effective voxel size of 3.05 x 3.05 x 3.05 $\mu m^3$. The tomography was produced by means of 2000 projections covering a total angle range of 180°. The acquisition time for each angular position was 300 ms. Data preprocessing, phase retrieval based on Paganin's algorithm, and tomographic reconstruction were performed with SYRMEP Tomo Project software and optimized scripts.

(2) The XPCT experiment at ANATOMIX beamline was performed with a filtered white beam peaked around 20 keV. The propagation distance was set at 0.2 m. The measurements were performed with an effective voxel size of 3.25 x 3.25 x 3.25 $\mu m^3$ and 0.65 x 0.65 x 0.65 $\mu m^3$, resulting from 2x and 10x optics respectively, coupled with Orca Flash 4.0 camera (sensor type CMOS, sensor array size 2048x2048, pixel size 6.5 mm 16-bit nominal dynamic range). The experiment was carried out recording 4000 projections in extended field-of-view (FOV) mode, an experimental procedure of acquisition which allows to almost double the effective horizontal width of the FOV of the detector. The rotation axis is moved close to either left or right side of the FOV and a dataset of projections, having size equal to the detector FOV, is collected over 360°. After properly stitching the sinograms, the reconstruction procedure can be performed as usual. Data pre-processing, phase retrieval based on Paganin's algorithm, and tomographic reconstruction were performed with PyHST software package.

(3) The experiment at I13-2 beamline used X-ray filtered-white beam peaked around 27 keV. The propagation distance was set at 0.10 m. Images were detected with a PCO.Edge 5.5 (sCMOS-technology, 2560 x 2160 pixels, 6.5 x 6.5 x 6.5 $\mu m^3$ voxel size and a 16-bit nominal dynamic range), coupled with a scintillator screen and 4x optics resulting in a total magnification of 8x due to the setup configuration. The effective voxel size was 0.8 x 0.8 x 0.8 $\mu m^3$. Tomography was performed acquiring 4001 projections, with an exposure time of 0.23 s per projection. The scans were performed in extended-FOV mode. Data pre-processing, phase retrieval based on Paganin's algorithm, and tomographic reconstruction were performed with SAVU, the tomographic data processing tool developed at Diamond Light Source Ltd.

(4) The XPCT experiment at SYRMEP, Elettra, was carried using pink beam with a mean energy of 23.5 keV. The sample-detector distance was set at 150 mm. The effective voxel size was 0.9 x 0.9 x 0.9 $\mu m^3$. The tomographic data were produced by means of 4000 projections covering a total angle range of 360°, in extended-FOV mode After properly stitching the sinograms, the reconstruction procedure can be performed as usual. Data preprocessing, phase retrieval based on Paganin's algorithm, and tomographic reconstruction were performed with SYRMEP Tomo Project software and optimized scripts.

The color scale of the tomographic images is a grey scale where the shades are proportional to electron density; white corresponds to the highest value of the density spectrum, whereas black corresponds to the lowest value hence, to features of lowest density.

**Image analysis**

Ring artifacts were removed by an improved frequency filtering [Massimi, L., et al., An improved ring removal procedure for in-line x-ray phase contrast tomography. Phys Med Biol, 2018. 63(4): p. 045007]. Image analysis was performed using ImageJ and the 3D rendering images were obtained using VGstudio Max software. To enhance the contrast and visualize structures developing in 3D and therefore lying on different slices, we exploited the z-projection of maximum and minimum intensities, which consists in projecting on the visualization plane the voxels of a set of continuous slices. Each pixel of the output image contains the maximum (or the minimum) value found along the axis perpendicular to that pixel.
The quantification of vascular damage - expressed as the number of vessels affected by blood barrier disruption – was performed through the following pipeline. Blood vessels and "clouds" were segmented in the input 3D grayscale image data using Weka Segmentation ImageJ plugin [Arganda-Carreras, I., et al., Trainable Weka Segmentation: a machine learning tool for microscopy pixel classification. Bioinformatics, 2017. 33(15): p. 2424-2426], which perform a semantic segmentation (SS) to distinguish different classes of objects in the same image. On the binarized images containing only the class of vessels, Skeletonize ImageJ plug-in [Arganda-Carreras, I., et al., (2010). Microsc.

Res. Tech. 73, 1019–1029] was applied to calculate the number vessels within the investigated volume of spinal cord. Clouds localization and quantification were instead performed through an automated counting process, using an ImageJ plug-in [Bolte, S. & Cordelies, F. P. (2006). J. Microsc. 224, 213–232]. The results obtained on vessels and clouds were matched to assess the percentage of damaged vessels with respect to the total number of vessels in the lumbar region of the spinal cord. To evaluate the accuracy of the semantic segmentation for the vascular network, a segmentation of the vessels through a 3D image segmentation process. Starting from the 3D tomographic image, an intensity threshold segmentation was performed to isolate the vascular network (appearing as white tubular objects). Then the obtained mask was compared with the result of SS.

To flatten the 3D surfaces of gut and eye, we exploited an algorithm that we recently developed, based on conformal mesh parameterization [Stabile, S., et al., A computational platform for the virtual unfolding of Herculaneum Papyri. Scientific Reports, 2021. 11(1): p. 1-11], and SheetMeshProjection Imagej plug-in [Wada, H. and S. Hayashi, Net, skin and flatten, ImageJ plugin tool for extracting surface profiles from curved 3D objects. microPublication biology, 2020], which use a coarsely spaced mesh to capture the surfaces of object. When the structure can be assimilated with good approximation to a regular geometric solid (such as a cylinder), we performed a standard surface unrolling procedure using Radial Reslice Imagej plug-in.

**ACKNOWLEDGMENTS**

This work was supported by Regione Puglia and CNR for Tecnopolo per la Medicina di Precisione. D.G.R. n. 2117 of 21.11.2018.